\documentclass[sigconf, nonacm]{acmart}
\usepackage{subcaption}

\usepackage{pdfrender}
\newcommand*{\boldcheckmark}{%
  \textpdfrender{
    TextRenderingMode=FillStroke,
    LineWidth=.5pt,
  }{\checkmark}
}

\AtBeginDocument{%
  \providecommand\BibTeX{{%
    \normalfont B\kern-0.5em{\scshape i\kern-0.25em b}\kern-0.8em\TeX}}}

\begin{document}

\title[STIXnet: A Novel and Modular Solution for Extracting All STIX Objects in CTI Reports]{STIXnet: A Novel and Modular Solution\\for Extracting All STIX Objects in CTI Reports}

\author{Francesco Marchiori}
\orcid{0000-0001-5282-0965}
\affiliation{%
 \institution{University of Padova}
 \streetaddress{Via Trieste, 63}
 \city{Padua}
 \country{Italy}
 \postcode{35121}
}
\email{francesco.marchiori@math.unipd.it}

\author{Mauro Conti}
\orcid{0000-0002-3612-1934} 
\affiliation{%
 \institution{University of Padova}
 \streetaddress{Via Trieste, 63}
 \city{Padua}
 \country{Italy}
 \postcode{35121}
}
\email{mauro.conti@unipd.it}

\author{Nino Vincenzo Verde}
\orcid{0000-0002-2379-6414}
\affiliation{%
 \institution{Leonardo S.p.A.}
 \streetaddress{Via Laurentina, 760}
 \city{Rome}
 \country{Italy}}
\email{nino.verde@leonardo.com}

\begin{abstract}
The automatic extraction of information from Cyber Threat Intelligence (CTI) reports is crucial in risk management. The increased frequency of the publications of these reports has led researchers to develop new systems for automatically recovering different types of entities and relations from textual data. Most state-of-the-art models leverage Natural Language Processing (NLP) techniques, which perform greatly in extracting a few types of entities at a time but cannot detect heterogeneous data or their relations. Furthermore, several paradigms, such as STIX, have become de facto standards in the CTI community and dictate a formal categorization of different entities and relations to enable organizations to share data consistently. 
\par
This paper presents STIXnet, the first solution for the automated extraction of all STIX entities and relationships in CTI reports. Through the use of NLP techniques and an interactive Knowledge Base (KB) of entities, our approach obtains F1 scores comparable to state-of-the-art models for entity extraction (0.916) and relation extraction (0.724) while considering significantly more types of entities and relations. Moreover, STIXnet constitutes a modular and extensible framework that manages and coordinates different modules to merge their contributions uniquely and exhaustively. With our approach, researchers and organizations can extend their Information Extraction (IE) capabilities by integrating the efforts of several techniques without needing to develop new tools from scratch.
\end{abstract}

\ccsdesc[500]{Security and privacy}
\ccsdesc[300]{Information systems~Information extraction}

\keywords{Cyber Threat Intelligence, Natural Language Processing, Information Extraction, STIX}

\maketitle

\section{Introduction}
\label{sec:introduction}

The increasing number of cyberattacks has concerned many people and organizations, which, now more than ever, are potentially exposing their data to breaches and attacks. In particular, companies must be aware of Advanced Persistent Threats (APTs), stealthy actors that establish a persistent presence in networks to steal information. APTs constitute one of the biggest concerns for most organizations since they are hard to identify and use complex attacks that are difficult to prevent~\cite{10.1007/978-3-662-44885-4_5}. Security experts use various types of intelligence to track their movements, motivations, and behavior to counter them. To do this, researchers analyze previous breaches to understand the techniques that APTs use, the vulnerabilities and malwares they exploit, and the tools used for their deployment~\cite{bianco2013pyramid}.

The collection and distribution of these kinds of data fall in the Cyber Threat Intelligence (CTI) field. Many researchers are involved in this discipline since, through this intelligence, companies can proactively defend against specific threat actors that might target them. In this way, organizations can protect specific assets more at risk than others and redirect their security budget in the best possible way~\cite{6657147}. For these reasons, companies must be up-to-date on the latest attacks, malwares, and techniques. To do so, they must collect intelligence by analyzing previous incidents or using external sources. Indeed, the distribution of this intelligence is provided by different vendors through reports and bulletins or by Open Source Intelligence (OSINT, i.e., any data that can be gathered for free). These reports are usually written in English and contain all the information on a particular incident or actor~\cite{WAGNER2019101589}.

The extraction of relevant information from these reports is usually performed by CTI analysts, trained to recognize the entities of interest and the relations between them. Once identified, intelligence must be annotated following certain standards to ensure machine readability and compatibility during distribution. One of the community's most popular and recognized standards is the Structured Threat Information Expression (STIX) language~\cite{barnum2012standardizing}, categorizing intelligence into different entities and relations types. However, this action can be time-consuming, given the reports' length and increased publication frequency in the last few years. To solve this issue, many attempts have been made to extract entities automatically and relations~\cite{You2022, https://doi.org/10.48550/arxiv.2004.14322, app9193945}. However, most state-of-the-art models focus on extracting a single or a few types of intelligence at a time, in which they obtain excellent results~\cite{10.1007/978-3-319-17040-4_24, 9023758}. In most real-world scenarios, the extraction of only a subset of entities and relations might not be enough to fully characterize the data contained in a report. Despite that, merging the results of different models is not trivial since conflicts might appear in the extracted data, and intelligence is often displayed through different standards and paradigms.

\textbf{Contribution.} In this paper, we present STIXnet, the first modular and extensible system for the automated extraction of all STIX entities and relationships in CTI reports. STIXnet works by leveraging different techniques, such as Natural Language Processing (NLP), to extract threat intelligence from the text of the report and identify the relevant pieces of information while also retrieving the relations among them. Our tool uses a rich Knowledge Base (KB) that contains CTI data from various sources and previous report extractions. The Knowledge Base can be enlarged with each execution and provide data for training Machine Learning (ML) and Deep Learning (DL) models used in some STIXnet modules. Through a graphical interface, the results of the STIXnet processing can be visualized as a graph in which nodes constitute entities and edges constitute relations. Each node can then be expanded with additional information stored in the database and thus provide a quick and interactive overview for each entity in the Knowledge Base. The architecture of our solution allows for the extraction of all entities and relations compliant with the STIX standard without running into structural constraints such as model retraining and dataset reannotation. In Table~\ref{tab:entities}, we show in more detail the types of entities that STIXnet can extract with respect to other models in the literature. It is worth noting that some of the models used for the comparison do not focus on STIX entities in particular, but their labels can be translated according to the STIX standard. Furthermore, in the comparison, we consider only entities that can be extracted from a report. Still, through our platform, it is possible to include entities compliant with the latest STIX 2.1 standard.

\begin{table*}[!htpb]
  \centering
  \caption{Comparison of the types of entities extracted by STIXnet.}
  \label{tab:entities}
  \begin{tabular}{lcccccccccccccc}
    \toprule
    \textbf{Model} & \rotatebox{90}{\textbf{Attack Pattern}} & \rotatebox{90}{\textbf{Campaign}} & \rotatebox{90}{\textbf{Course of Action}} & \rotatebox{90}{\textbf{Identity}} & \rotatebox{90}{\textbf{Indicator}} & \rotatebox{90}{\textbf{Infrastructure}} & \rotatebox{90}{\textbf{Intrusion Set}} & \rotatebox{90}{\textbf{Location}} & \rotatebox{90}{\textbf{Malware}} & \rotatebox{90}{\textbf{Malware Analysis}} & \rotatebox{90}{\textbf{Report}} & \rotatebox{90}{\textbf{Threat Actor}} & \rotatebox{90}{\textbf{Tool}} & \rotatebox{90}{\textbf{Vulnerability}}\\
    \midrule
    Weerawardhana et al. ~\cite{10.1007/978-3-319-17040-4_24} & & & & & \checkmark & & & & & & & \checkmark & \checkmark & \checkmark \\
    Li et al.~\cite{9023758} & \checkmark &  &  &  &  &  &  &  & \checkmark &  &  & \checkmark & \checkmark &  \\
    Zhou (2022) et al.~\cite{Zhou2022} &  & \checkmark &  & \checkmark &  &  &  & \checkmark &  &  &  & \checkmark & \checkmark &  \\
    Zhou (2023) et al.~\cite{zhou2023cdtier} &  & \checkmark &  & \checkmark &  &  &  & \checkmark &  &  &  & \checkmark & \checkmark & \\
    Ranade et al.~\cite{9671824} & \checkmark &  & \checkmark &  &  &  &  &  & \checkmark &  &  & \checkmark & \checkmark & \checkmark \\
    Wang et al.~\cite{wang2022cyber} & \checkmark & &  & \checkmark & \checkmark &  &  & \checkmark & \checkmark &  &  & \checkmark & \checkmark & \checkmark \\
    \textbf{STIXnet} & \boldcheckmark & \boldcheckmark & \boldcheckmark & \boldcheckmark & \boldcheckmark & \boldcheckmark & \boldcheckmark & \boldcheckmark & \boldcheckmark & \boldcheckmark & \boldcheckmark & \boldcheckmark & \boldcheckmark & \boldcheckmark \\
    \bottomrule
\end{tabular}
\end{table*}

The main contributions of our work can be summarized as follows:
\begin{itemize}
    \item We propose the first system for automatically extracting \textbf{all} types of STIX entities (18) and relations (more than 100).
    \item We propose a novel framework for managing and coordinating several modules for Information Extraction.
    \item We propose a methodology for integrating results from different modules through a confidence value and with minimal supervision.
    \item We make our testbed (dataset and annotated reports through LabelStudio\footnote{\url{https://labelstud.io/}}) and the code for some modules available at \url{https://anonymous.4open.science/r/STIXnet-7710}.
\end{itemize}

\textbf{Organization.} The rest of the paper is organized as follows. In Section~\ref{sec:background}, we present some background on Natural Language Processing techniques and Cyber Threat Intelligence. Section~\ref{sec:relatedworks} presents a review of previous works on Information Extraction (IE) from natural language reports in the field of CTI. The proposed methodology and the pipeline are presented in Section~\ref{sec:methodology}, followed by a formal analysis of the results in Section~\ref{sec:results}. Finally, Section~\ref{sec:conclusions} concludes this work.
\section{Background}
\label{sec:background}

We now give a more thorough background on the techniques that we use in the methodology (Section~\ref{subsec:nlp}) and further expose the challenges of Cyber Threat Intelligence (Section~\ref{subsec:cti}).

\subsection{Natural Language Processing}
\label{subsec:nlp}

Given the fluency and convenience of natural language for human interactions, the field of Natural Language Processing is born to develop algorithms and models that can comprehend and analyze this type of language. These algorithms can also include Machine Learning or Deep Learning techniques, which create many opportunities for its applications. With ML and DL techniques, many documents can be automatically processed to extract named entities, detect their attributes, and retrieve their existing relations. For this reason, NLP has become particularly suitable for tasks involving the analysis of multiple reports and extracting a predefined set of data in a text~\cite{Chowdhary2020}.

NLP can be applied in every domain in which human language is the main vector through which information is conveyed, e.g., speech recognition (i.e., speech-to-text, the act of translating voice data into text data), Part-Of-Speech tagging (i.e., grammatical tagging or POS tagging, the act of determining the part of speech of a particular word based on its context) and Named Entity Recognition (i.e., NER, the act of identifying specific words in a text as a specific type of entity)~\cite{8629225}. In particular, this last technique is heavily used for Information Extraction tasks where a large number of text data is involved~\cite{Piskorski2013}, such as medical applications~\cite{Weegar2021}, scientific research~\cite{10.1007/978-3-030-00671-6_8} and cybersecurity intelligence~\cite{You2022}.

Information Extraction also comprises Relation Extraction, i.e., retrieving and classifying the semantic relationships between two (or more) tokens inside a text~\cite{10.1007/978-3-319-12580-0_2}. This task can become particularly important in sequence with the Entity Extraction task. In this way, the information in textual data becomes intertwined, and a knowledge graph of the processed text can be created~\cite{https://doi.org/10.48550/arxiv.2106.00459}.

\subsection{Cyber Threat Intelligence}
\label{subsec:cti}

According to one of the definitions of the Computer Security Research Center at NIST\footnote{\url{https://csrc.nist.gov/glossary/term/cyber\_threat}}, a cyber threat can be defined \textit{"any circumstance or event with the potential to adversely impact organizational operations [...] via unauthorized access, destruction, disclosure, modification of information, and/or denial of service"}. CTI is the field that studies these threats and analyzes the intentions of the threat actors, the techniques they use, and the tools and malwares they deploy. In this way, it is possible to profile the activities of the malicious actors and thus design more effective cyber defense strategies~\cite{WAGNER2019101589}.

Several types of intelligence are treated by CTI, which can be more or less technical depending on the target audience that must digest it. Also, each piece of intel follows a specific life cycle from planning and direction to its dissemination and integration~\cite{porkorny2018phases}.

To efficiently share and distribute the collected information, a common protocol must be in place to avoid misunderstandings between the teams that consume it. To address these problems, the STIX (Structured Threat Information eXpression) standard has been created~\cite{barnum2012standardizing}. This standardized language has been created by MITRE\footnote{\url{https://www.mitre.org/about/corporate-overview}} and is driven by the collaboration of many individuals who keep it up-to-date. Including different types of entities and relationships makes it possible to accurately represent the information in a cyber security report through a STIX file. The latest version of the software is STIX 2.1\footnote{\url{https://docs.oasis-open.org/cti/stix/v2.1/csprd01/stix-v2.1-csprd01.html}}, which includes 18 STIX Domain Objects (SDOs) and more than 100 possible relations among them.

While STIX provides a standardized language for disseminating CTI, it does not provide intelligence. One of the most popular frameworks for Cyber Threat Intelligence is the MITRE ATT\&CK framework, a publicly accessible Knowledge Base of TTPs extracted from real-world CTI reports that can be used as a foundation for building a personalized database of threat intelligence~\cite{strom2018mitre}. Entities in the ATT\&CK Knowledge Base are categorized with different labels, of which the ones of interest for the scope of this paper are \textit{tactics}, \textit{techniques}, \textit{groups}, and \textit{software}. However, since these entities do not have an official one-to-one correspondence with the STIX entity types, a mapping is needed for its referencing, which can be found in Table~\ref{tab:mitre2stix}.

\begin{table}[!htpb]
  \centering
  \caption{Conversion of ATT\&CK entity types to STIX Objects.}
  \label{tab:mitre2stix}
  \begin{tabular}{ll}
    \toprule
    \textbf{ATT\&CK Entity Type} & \textbf{STIX Object}\\
    \midrule
    Tactic & \texttt{x-mitre-tactic}\\
    Technique & Attack Pattern\\
    Mitigation & Course of Action \\
    Group & Intrusion Set \\
    Software & Malware/Tool \\
    \bottomrule
\end{tabular}
\end{table}
\section{Related Works}
\label{sec:relatedworks}

The need for the automatic processing of CTI reports and retrieving entities and relations has pushed researchers to adopt Information Extraction methods in the field. However, this is not an easy task due to many reasons. First, information in raw text reports can be conveyed differently through semantics, and bulletin styles can differ from vendor to vendor. Furthermore, reports might (and frequently do) include new entities, making the usage of a static and non-interactive database of entity templates inefficient. Moreover, ever-changing Indicators of Compromise (IOCs, e.g., IP addresses, hashes, URLs, Bitcoin addresses) must be recognized and linked to their respective actor or malware. For these reasons, many different models have been proposed to push research on constructing a Knowledge Graph in Cybersecurity~\cite{9480953}. One of the aspects that can be noticed in the current literature on IE techniques applied to CTI reports is that many of the proposed models focus on one or a few types of entities/relations at a time while neglecting the others. This allows researchers to obtain impressive results in a narrow domain that does not always reflect the whole needs of the CTI community. Moreover, while being very efficient, using Machine Learning and Deep Learning models in Information Extraction has a few drawbacks. First, it is not easily scalable on wider domains to include more entity types and thus requires fully retraining the model while making changes to the architecture. Secondly, the training dataset might be annotated with just a few entity types, and to include other STIX types, analysts should perform the procedure all over again. This is quite time-consuming and can become very costly for companies and organizations.

In You et al.~\cite{You2022}, researchers have developed a model that can retrieve Tactics, Techniques, and Procedures (TTPs) with an accuracy of 0.941, which constitutes the state-of-the-art for this task. However, not only do TTPs not reflect the overall spectrum of Cybersecurity entities that should be extracted in a report, but the number of these TTPs is just 6. At the same time, the MITRE ATT\&CK Knowledge Base indicates at least 14 tactics and 191 techniques (just in enterprise environments and thus neglecting mobile and ICS attacks).

Similar work has been previously done by Legoy et al. with rcATT, a Python tool used to predict MITRE ATT\&CK tactics and techniques from cyber threat reports~\cite{https://doi.org/10.48550/arxiv.2004.14322}. It has a maximum precision of around 0.75 in both tactics and techniques. These reduced performance levels are justified by an increased number of labels, comprising 215 MITRE ATT\&CK techniques and 12 tactics. However, this work was published in 2020, and the MITRE ATT\&CK framework has changed with new techniques and tactics, so careful reparametrization is needed, and (as stated in the future work's sections of the paper) retraining it on a bigger dataset might improve its performance.

A more general approach that can tackle a broader domain of entities is the work of Gasmi et al.~\cite{app9193945}, which includes both entity extraction and relation extraction on data from the National Vulnerability Database (NVD)\footnote{\url{https://nvd.nist.gov/}}. Researchers addressed seven common entity types and six relationship types. This database contains CVE items, i.e., disclosed cybersecurity vulnerabilities specifically formatted to be more easily cataloged, evaluated, and shared among the community. The tool reaches a precision value of 89\% on the entity extraction task and 92\% on the relation extraction task. However, the data used for training and testing consists of vulnerability descriptions that, while written in natural language, present a common structure and thus do not reflect the variance that might be present in more common CTI reports or bulletins.

While some proposed works leverage Machine Learning models, which perform particularly well in narrow domains, other string tagging techniques can tackle the entity extraction task.

\begin{itemize}
    \item \textbf{Name-Matching Strings}: if a Knowledge Base containing the entity names is already in place inside the platform, it can match words inside a text. Even though the construction of the KB can be time-consuming, there are a lot of public sources from which to retrieve information on these entities, and it is also possible to provide aliases for each of them, thus being able to recognize pseudonyms and still link them to the correct entity.
    \item \textbf{RE-Matching Strings}: Indicators of Compromises can be found by their particular structure constant across the same type of IOC. For example, IP(v4) addresses are always be written in the format \texttt{XXX.XXX.XXX.XXX}, i.e., four sets of numbers from 0 to 255 separated by a dot character. Different rules apply to different types of entities. Still, in the domain of words with distinct character structures, it is possible to use regular expression tools to identify and extract them.
    \item \textbf{Verb-Related String}: this technique is used when the other two fail in extracting entities like companies and new malware names, which do not present any particular character structure and might not be present inside the Knowledge Base. These entities can be retrieved by analyzing each sentence through POS Tagging and Dependency Parsing. After that, we can retrieve the verb to match it with a predefined set of words that might indicate the presence of an entity.
\end{itemize}

STIXnet expands these results and further generalizes the entities and relationships that can be extracted in a document to comprise all STIX entities and relationships, which most closely represent the information an analyst should extract from CTI reports.
\section{Methodology}
\label{sec:methodology}

In this section, we present STIXnet in more detail on how it works and which algorithms and models are used to perform Information Extraction on unstructured Cyber Threat Intelligence reports. We first show its pipeline and briefly overview the various modules (Section~\ref{subsec:pipeline}). Then we give more details on its main components: the text extraction module (Section~\ref{subsec:textextraction}), the entity extraction module (Section~\ref{subsec:entityextraction_methodology}) and the relation extraction module (Section~\ref{subsec:relationextraction_methodology}).

\subsection{Pipeline}
\label{subsec:pipeline}

\begin{figure*}[!htpb]
  \centering
  \includegraphics[width=\linewidth]{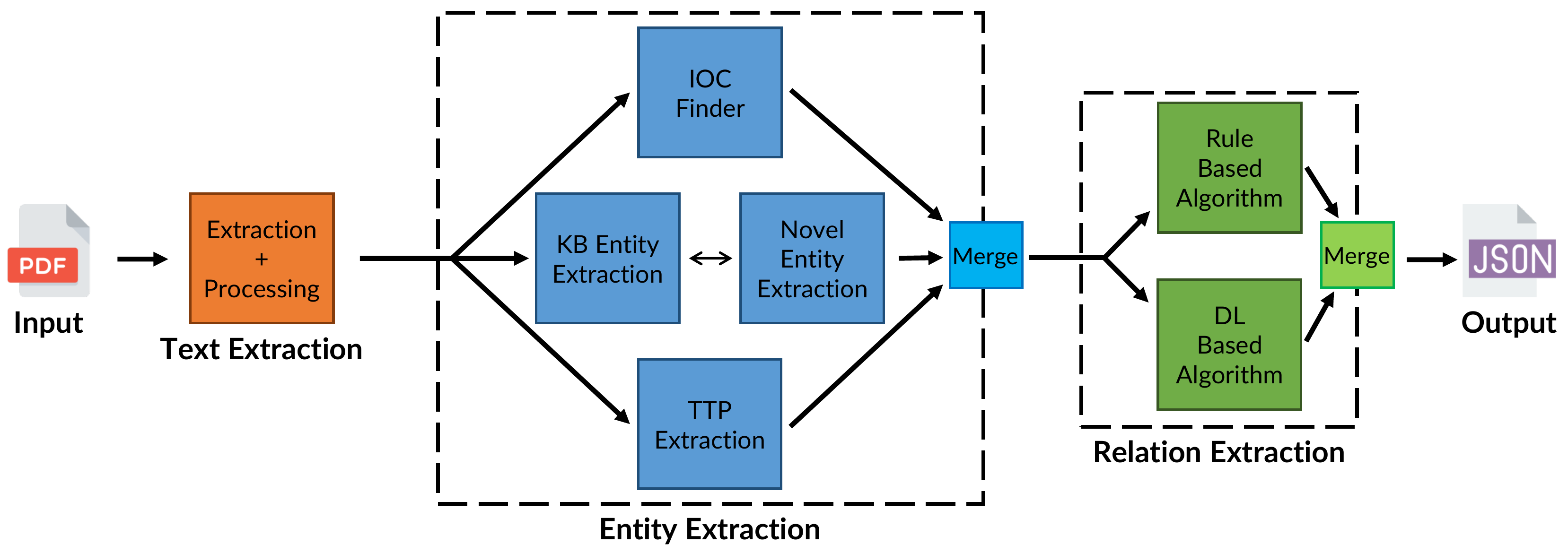}
  \caption{Pipeline of the STIXnet platform.}
  \label{fig:pipeline}
\end{figure*}

STIXnet performs a highly complex Information Extraction task on many different types of entities and relations. There are indeed 18 different types of entities compliant with the STIX standard and more than 100 types of relations. To accomplish this, STIXnet uses different modules for each one of the different tasks that it must achieve: textual extraction, entity extraction, and relation extraction. A graphic overview of the STIXnet platform is shown in Figure~\ref{fig:pipeline}.

\begin{itemize}
    \item \textbf{Text Extraction}: the first module converts the program's input into raw text. While doing this, artifacts are inevitably be introduced in the text, and therefore they must be handled to obtain a single string of characters in a common encoding.
    \item \textbf{Entity Extraction}: this module handles the extraction of the different entities in the text. It uses four different sub-modules to accomplish that:
    \begin{itemize}
        \item \textbf{IOC Finder}: we extract Indicators Of Compromise by using different regular expression rules and looking at text patterns.
        \item \textbf{Knowledge Base Entities Extraction}: using the entities in a Knowledge Base, we use an efficient algorithm for string search in a text to retrieve names and aliases. We mitigate errors and false positives caused by this approach using NLP techniques.
        \item \textbf{Novel Entities Extraction}: using NLP libraries, we extract entities not present in the Knowledge Base, which can then be integrated into the KB.
        \item \textbf{TTPs Extraction}: techniques and tactics are not always represented in a text by their name but can be implicit and not explicitly expressed. We use a Machine Learning model trained on MITRE ATT\&CK tactics and techniques to recognize them.
    \end{itemize}
    \item \textbf{Relation Extraction}: from the extracted entities, we now retrieve the relations. We use two sub-modules for this task:
    \begin{itemize}
        \item \textbf{Rule-Based Approach}: using NLP techniques, we perform Dependency Parsing and compute the shortest paths between entities. Comparing the verb inside the path with the ones in the STIX relationships, we estimate the most similar one with a degree of confidence.
        \item \textbf{Deep Learning Based Approach}: to adjust the results of the previous approach, we also compute embeddings from the sentences with a Deep Learning model. We then determine the similarity between these embeddings and those computed from the list of relationship labels.
    \end{itemize}
    \item \textbf{Output}: we create a JSON file from the extracted entities and relations, which is processed by the graphical interface of the platform to be interactive and dynamic.
\end{itemize}

Eventual interactions with the Knowledge Base or other platform components are disclosed in the individual sections of the different modules. Indeed, we show that by interacting with the database, it is possible to improve performance over time, particularly if an analyst decides to validate the results of the STIXnet output. 

Furthermore, the structure of the STIXnet pipeline allows for easy and immediate management of the different modules. Formally defining the interactions between modules and submodules allows results from different pipeline components to be compared and merged in a unique output. Thus, researchers and organizations can implement the STIXnet framework in different IE scenarios and add or remove components depending on their needs.

\subsection{Text Extraction}
\label{subsec:textextraction}

One of the platform's most relevant and sensible aspects is its input. As mentioned, STIXnet can take in input various reports and bulletins, which can come from various vendors or sources. For this reason, reports have some stylistic and linguistic differences. However, data must be converted univocally before processing the raw text to have consistent processing between different inputs and a common ground for evaluating the various modules. This means considering many different aspects that can change from input to input. First of all, we must be able to parse text from files with different formats, and thus we use Apache Tika\footnote{\url{https://tika.apache.org/}} for \texttt{pdf} and \texttt{doc} files and ConvertAPI\footnote{\url{https://www.convertapi.com/}} to extract text from the HTML data of web reports. However, since text can be formatted in many different ways, we must process it to remove artifacts, fix line breaks, and remove eventual sanifications on IP and other addresses. This phase is crucial to ensure that the following modules are presented with a clean input; otherwise, artifacts are propagated in the pipeline and compromise the results.

\subsection{Entity Extraction}
\label{subsec:entityextraction_methodology}

This section tackles the different submodules used for entity extraction: IOC Finder (Section~\ref{subsub:iocfinder}), a rule-based entity extractor (Section~\ref{subsub:kbentext}), a novel entity extractor (Section~\ref{subsub:novelentityextraction}), and a TTPs extractor (Section~\ref{subsub:ttpext}). Finally, we clarify how the submodules interact with one another and merge their results (Section~\ref{subsub:subint}).

\subsubsection{IOC Finder}
\label{subsub:iocfinder}

The particular structure of Indicators Of Compromise allows us to use regular expression (Regex) rules to find them in the text. Moreover, in some of the reports distributed by CTI vendors, the end of the document is often presented with a table containing the IOCs of interest for that particular topic. While different, all these indicators share a common structure across each type, which can be recognized with Regex rules without applying NLP techniques. Some examples of the types of IOCs supported by IOC Finder can be found in Table~\ref{tab:iocfinder}. During the execution of STIXnet and after processing the report as raw text, the first step is to run the IOC Finder submodule on it, which returns a dictionary containing the entities found.

We implement this module by forking an open-source project by Floyd Hightower\footnote{\url{https://github.com/fhightower/ioc-finder}}. To adapt it for our pipeline, we contributed to the main project by updating some libraries to a newer version and adding the ability to track the position of the found IOCs. The code of the forked project can be found in our repository.

\begin{table}[!htpb]
  \centering
  \caption{Examples of IOCs and their structure.}
  \label{tab:iocfinder}
  \begin{tabular}{ll}
    \toprule
    \textbf{IOC Type} & \textbf{IOC Structure}\\
    \midrule
    ATT\&CK Techniques & \texttt{T1518} or \texttt{T1518.001}\\
    CVEs & \texttt{CVE-2021-44228} \\
    Email Addresses & \texttt{example@mail.com} \\
    File Paths & \texttt{/path/to/file} \\
    MD5s & \texttt{e802c6b77dd5842906ed96ab1674c525} \\
    \bottomrule
\end{tabular}
\end{table}

\subsubsection{Knowledge Base Entity Extraction}
\label{subsub:kbentext}

After finding IOCs, all other entities of interest do not share a common structure and thus cannot be found through regular expression rules. Thus, we leverage a rich Knowledge Base integrated with multiple OSINT that explicitly indicate which names represent important entities and allow us to link them to their correct entity type. For this reason, a rule-based algorithm can search specific words in the text, retrieve their position and thus highlight them as extracted entities. The different sources for the intelligence are:

\begin{itemize}
    \item \textbf{Knowledge Base}: Leonardo S.p.A., an Italian multinational company that collaborated in this research, provided a rich database of STIX entities. Cyber threat intelligence analysts built this database over the years at their Security Operation Center (SOC), which read and manually annotated entities from many reports.
    \item \textbf{MITRE ATT\&CK}: the ATT\&CK framework can be used as a source of intelligence for different entities such as techniques, tactics, groups, and software (of which the conversion to the STIX standard has been disclosed in Table~\ref{tab:mitre2stix}). To retrieve this data, we use the Trusted Automated Exchange of Intelligence Information (TAXII) application protocol~\cite{connolly2014trusted}, which allows for exchanging threat intelligence over HTTPS and defines a RESTful API that can be used to provide or collect data.
    \item \textbf{Locations}: to retrieve the names of countries and continents, we used a \texttt{csv} file in which each nation is associated with its nationality. In this way, we are able to identify locations even when used as an attribute to another entity (e.g., "a \textit{Russian} malware").
\end{itemize}

After retrieving the entities, a quick pre-processing is performed to unify their formats and add the possibility of aliases for each one. Through aliases, it is possible to recognize an entity in a text and map it to the correct one, avoiding duplicates and fixing the issue of multiple names for a single Advanced Persistent Threat. We then use the Aho-Corasick algorithm to find terms of this thesaurus of words in the report~\cite{10.1145/360825.360855}. To mitigate false positives, we process each sentence with NLP techniques, in particular, Part-Of-Speech Tagging (POS tagging), allowing us to assign part-of-speech tags to each word (e.g., noun, verb). After defining a table of entity types and their possible POS tags, we use it to compare the entities found in the report with their extracted POS tag. For example, "us" can be used as a pronoun or can be used as a noun to reference the United States, constituting an entity of type "location".

\subsubsection{Novel Entities Extraction}
\label{subsub:novelentityextraction}

\begin{figure*}[!htpb]
  \centering
  \includegraphics[width=\linewidth]{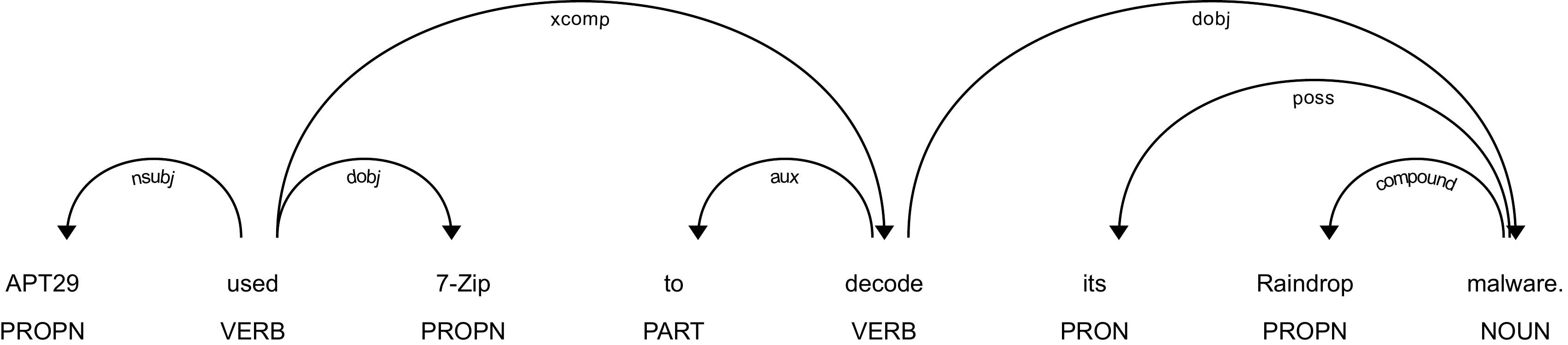}
  \caption{Example of dependency graph generated from a sentence.}
  \label{fig:spacygraph}
\end{figure*}

Some CTI reports are published to spread awareness of newly discovered actors, malwares, or techniques. These entities are thus named by CTI researchers and analysts and, for this reason, are most likely not present in the Knowledge Base. To find these new entities, we can leverage the previous execution of POS Tagging to extend its results and create a dependency graph from the tokens found in each sentence. In this way, we identify specific patterns used in the text to express a new entity. To create such a graph, we leverage both the POS tags and the dependencies between the tokens, as shown in Figure~\ref{fig:spacygraph}. To perform this processing, we use Spacy\footnote{\url{https://spacy.io/}}, a free, open-source library for advanced NLP in Python. By looking at numerous reports and bulletins from different vendors and sources, we can identify a limited number of ways a new entity can be introduced, allowing us to write pattern rules that the NLP processing can recognize.

\subsubsection{TTPs Extraction}
\label{subsub:ttpext}

While malwares and threat actors are often explicitly mentioned, some other entities are not and can be referenced without categorically stating their names. It is the case of tactics and techniques, which constitute the TTPs mapped into the STIX objects "x-mitre-tactic" and "attack pattern". Rule-based methods cannot be used to retrieve these entities since the variance that characterizes the expression of these concepts is too broad and is hardly definable through a set of rules. For this reason, a multi-label text classification model must be deployed and trained on the MITRE ATT\&CK Knowledge Base of tactics and techniques. We already presented a tool named rcATT~\cite{https://doi.org/10.48550/arxiv.2004.14322} that suffered from its age since it was published in 2020, and the MITRE ATT\&CK framework has had several changes and renovations ever since. The source code for rcATT is publicly available in their GitHub repository\footnote{\url{https://github.com/vlegoy/rcATT}}, and thus we retrained it from scratch with new techniques and tactics. Also, to address one of the limitations and future works of the paper presenting rcATT, we expanded the training dataset with more data from MITRE ATT\&CK descriptions and external sources for each tactic and technique. The new dataset, the pre-trained models, and the updated code for its training can be found in our repository.

\subsubsection{Submodules Interaction}
\label{subsub:subint}

As stated in Section~\ref{subsec:pipeline}, the pipeline structure of our framework allows us to differentiate the tasks in many modules. For entity extraction, in particular, we identified four different submodules independent from one another. Their input is always the same and is constituted by the textual data of the report. Once each submodule produces an output, a final check is performed to ensure no overlapping occurs during the processing. This process includes cross-checking the found entities in the Knowledge Base to maximize their confidence in their extracted type and the addition of the novel entities. More details are disclosed in Section~\ref{subsec:entityextraction_results}. Finally, all the entities are merged into one data structure, constituting the following module's input. Since this procedure is automatic, submodules can be added and removed in the pipeline according to the user's needs, and no conflicts occur during the process.

\subsection{Relation Extraction}
\label{subsec:relationextraction_methodology}

This module can retrieve relations between the found entities while processing the sentences in the raw text. However, one of Spacy's limitations is its inability to grasp relations between distant entities in a text. To address this, we propose two different approaches for relation extraction. In the first approach, we leverage POS Tagging and Dependency Parsing to compute a graph of each sentence and retrieve relations by looking at the shortest paths between entities (Section~\ref{subsub:rulebased}). In the second approach, we use a Transformer model to compute embeddings of the sentences and compute their similarity (Section~\ref{subsub:dlbased}). Finally, we clarify how the submodules interact with one another and merge their results (Section~\ref{subsub:relsubint}).

\subsubsection{Rule Based Approach}
\label{subsub:rulebased}

The main idea of the rule-based approach is to leverage the dependency graphs already computed by Spacy for novel entity extraction (Section~\ref{subsub:novelentityextraction}) and use graph theory functions to grasp the relation between two entities inside a sentence. In particular, we can process the dependency graph and retrieve the relations between any couple of nodes by discovering the Shortest Dependency Path (SDP), i.e., the shortest path between two nodes in the graph. It has been observed in other studies that the nodes in the SDPs usually contain the necessary information to identify a relationship between two entities while also being dependent on the structure and semantic complexity of the sentence~\cite{hua2016shortest, xu-etal-2015-classifying}.

After retrieving the shortest path between each couple of entities in the sentence, we extract their STIX type and focus on the verbs in the path. For example, considering the sentence in Figure~\ref{fig:spacygraph}, the extracted paths (between the three entities \texttt{"APT29"}, \texttt{"7-Zip"}, and \texttt{"Raindrop"}) are:
\begin{itemize}
    \item \texttt{[APT29, \underline{used}, 7-Zip]};
    \item \texttt{[APT29, \underline{used}, \underline{decode}, malware, Raindrop]};
    \item \texttt{[7-Zip, \underline{used}, \underline{decode}, malware, Raindrop]}.
\end{itemize}
By looking at the entity types and the root form of the verbs (underlined in the previous example), it is possible to compare them with the ones found in the list of STIX relationships and thus label the path as a specific STIX Relationship Object (SRO). However, since the verbs used to describe the relationship in the sentence might not be the same as the associated SRO, we use a similarity function to determine their likeliness. If it surpasses a certain threshold, we can consider them synonyms. To accomplish this, we use the Wu \& Palmer similarity function~\cite{https://doi.org/10.48550/arxiv.cmp-lg/9406033}, which, given two words and their synsets (i.e., groupings of similar words that express the same concept), outputs a value in the range $\left[0,1\right]$, where $1$ means maximum similarity. For this reason, this value can be used as a confidence measure of the relation. The taxonomy used for this task is the WordNet taxonomy, an extensive lexical database of English words developed by Princeton University~\cite{Fellbaum2010}.

For example, considering the SDP \texttt{[APT29, used, 7-Zip]}, the entity extraction module identified \texttt{"APT29"} as an \texttt{intrusion-set} and \texttt{"7-Zip"} as a \texttt{tool}. In the list of SROs, there is only one entry containing both an \texttt{intrusion-set} and a \texttt{tool}, which is "intrusion-set uses tool": since the root form of the verb in the SDP is equal to the one in the SRO, we label it accordingly with maximum confidence. Instead, considering another SDP \texttt{[APT29, attacks, the, US]} (where \texttt{"US"} is identified as a \texttt{location} entity), there are two SROs including both an \texttt{intrusion-set} and a \texttt{location}: "intrusion-set originates-from location" and "intrusion-set targets location". In this case, none of the root forms of SROs verbs ("originate" and "target") coincide with the one in the SDP ("attack"), and thus we compute the similarity between them:
\begin{equation*}
    wup("attack", "originate") = 0.4,
\end{equation*}
\begin{equation*}
    wup("attack", "target") = 0.5.
\end{equation*}
While keeping the confidence threshold at 0.5, we can label the SDP as a relationship of type "intrusion-set targets location".

The confidence value in the extracted relations becomes particularly important when dealing with sentences containing multiple entities. Indeed, the number of relations increases exponentially with the number of entities found, and some of the extracted relations could not exist. To include a "non-relation" label in this task, we set a threshold for the confidence, under which we discard the extracted relations.

\subsubsection{Deep Learning Based Approach}
\label{subsub:dlbased}

The rule-based approach is particularly efficient when dealing with simple phrases or when entities are close in the graph. However, many elements might be introduced in the SDP whenever two entities are far from each other. The verb with the highest similarity could be linked to different tokens in the text and might not reach the confidence threshold. To address this problem, we use embeddings, fixed-size vectors that can also be generated from textual data by Deep Learning models such as Transformers~\cite{https://doi.org/10.48550/arxiv.1706.03762}.

In this specific case of relation extraction, we are interested in computing the similarity between each sentence's embeddings and the STIX relationships' embeddings. To perform these embeddings, the best tool at our disposal is Sentence BERT (SBERT)\footnote{\url{https://www.sbert.net/}}, a variation of the BERT model (Bidirectional Encoder Representations from Transformers) developed by Google AI language~\cite{https://doi.org/10.48550/arxiv.1905.05950}. While BERT constitutes the state-of-the-art in many NLP applications, it becomes inefficient when dealing with a large corpus of sentence processing. SBERT addresses this problem using siamese and triplet network structures, drastically reducing processing time~\cite{reimers-2019-sentence-bert}.

For its STIXnet implementation, we compute the embeddings of the different STIX relationships and the sentences extracted from the report. For each sentence, we perform a pre-processing procedure for each contained entity couple by substituting their tokens with their extracted STIX type. Then, we compute the cosine similarity between these embeddings and normalize it to use it as a confidence value. We also use a threshold for the confidence of the relation to discriminate false positives (0.5).

\subsubsection{Submodules Interaction}
\label{subsub:relsubint}

As with the entity extraction module, the relation extraction task is divided into different submodules that work independently. While their input is always the same (i.e., textual data from the report and the previously identified entities), the two submodules perform virtually the same task this time. Thus we expect a degree of overlap in their results. However, their coexistence is necessary to extract relations from both simple and complex scenarios. To handle conflicts in the possible output, we use the confidence values generated during the processing and keep the relations between entities with maximum confidence, which must also be over the acceptance threshold. Therefore, users can add their desired submodules for the relation extraction task, and STIXnet will automatically merge their results.
\section{Evaluation}
\label{sec:results}

We formally evaluate the proposed entity and relation extraction modules in STIXnet. We use three metrics to evaluate the modules: precision, recall, and F1-score. We define True Positives (TP), False Positives (FP), and False Negatives (FN) as follows.

\begin{itemize}
    \item \textbf{True Positives}: entities or relations correctly classified by the model.
    \item \textbf{False Positives}: entities and relations found by the model but misclassified or that do not constitute an entity or a relation.
    \item \textbf{False Negatives}: entities and relations not found by the model but present in the report.
\end{itemize}

The metrics for the evaluation are Precision, Recall, and F1 Score and are defined as follows.

\begin{equation}
    Precision = \frac{TP}{TP + FP},
\end{equation}

\begin{equation}
    Recall = \frac{TP}{TP + FN},
\end{equation}

\begin{equation}
    F1 = 2\frac{Precision \cdot Recall}{Precision + Recall}.
\end{equation}

Given the lack of available annotated reports in the literature, we generated our own dataset of CTI reports to evaluate our model. Each report treats a group or threat actor from the MITRE ATT\&CK framework. All their related data has been extracted from their official descriptions and the external sources listed by the ATT\&CK APIs. We then manually label each report with LabelStudio, a free and open-source software for data labeling. Both the dataset and the annotations are free to use and accessible in our repository. Annotations are exported in a \texttt{JSON} file and can be graphically visualized through the LabelStudio software.

To show the effectiveness of both entity and relation extraction, we tackle the evaluation of the modules separately, respectively, in Section~\ref{subsec:entityextraction_results} and Section~\ref{subsub:relationextraction_results}.

\subsection{Entity Extraction}
\label{subsec:entityextraction_results}

While evaluating the entity extraction module for STIXnet, we must consider the different sub-modules separately since they perform different independent operations.

To have a baseline for comparing the results, we first evaluate the completeness of the deployed Knowledge Base in a fixed position in time, i.e., not being enhanced by adding new entities when found. We take the whole dataset and run the rule-based entity extraction sub-module on each report. We then extract precision, recall, and F1 scores and compute the mean of these scores over the number of reports processed. We compare this approach with other rule-based algorithms for entity extraction found in the literature. Since rule-based approaches are domain-dependent and language-dependent, in Table~\ref{tab:kbeval} we compare our approach with other works in literature that operate in specific domains to more accurately resemble our task (for~\cite{10.1093/jamia/ocz109}, we extracted the evaluation of the hybrid model since it more accurately resembles our rule-based algorithm). While the compared models are not designed for CTI data extraction, to the best of our knowledge, no other tools in the literature perform such an evaluation on rule-based approaches for STIX entities. As a note, the ground truth constituted by the manually annotated reports also contains novel entities and TTPs (not extractable with just this submodule), thus highlighting the contribution of the sub-module in the whole entity extraction system.

\begin{table*}[!htpb]
  \centering
  \caption{Baseline evaluations for the Knowledge Base entity extraction sub-module and comparisons.}
  \label{tab:kbeval}
  \begin{tabular}{llllll}
    \toprule
    \textbf{Model} & \textbf{Domain} & \textbf{Entities} & \textbf{Precision} & \textbf{Recall} & \textbf{F1 Score}\\
    \midrule
    Godény~\cite{6406529} & Consumer electronics & Product names & N/A & N/A & 0.221 \\
    Quimbaya et al.~\cite{QUIMBAYA201655} & Electronic health records & Diagnosis, treatment & 0.630 & 0.573 & 0.600 \\
    Chen et al.~\cite{10.1093/jamia/ocz109} & Clinical trial cohort selection & Patient data and conditions & N/A & N/A & 0.845 \\
    \textbf{STIXnet Rule-Based Algorithm} & \textbf{CTI Reports} & \textbf{All STIX entity types} & \textbf{0.835} & \textbf{0.869} & \textbf{0.846}\\
    \bottomrule
\end{tabular}
\end{table*}

We then evaluate the novel entity extraction sub-module separately. We created a new dataset of sentences containing non-existent entities with made-up names for its evaluation. We do this since, in the original dataset, some reports do not contain any new entities and would thus bias the results of this evaluation. By creating new sentences instead, we ensure a constant number of novel entities that the sub-module can extract. In such a scenario, the isolated sub-module reaches a precision of 0.927, a recall of 0.854, and an F1 score of 0.889. Given the system's modularity and the submodules' specific task, a comparison with other models for Information Extraction in Cyber Threat Intelligence is given in Table~\ref{tab:noveleval} (the results of STIXnet extraction are obtained with the dynamically augmented Knowledge Base). Results are obtained after combining the contribution of the different approaches and evaluating them in the same dataset used for the baseline evaluation. Also, while focusing on CTI applications, we highlight the number of entity types the compared models can extract.

\begin{table*}[!htpb]
  \centering
  \caption{Comparison of evaluations for the entity extraction task in the CTI domain.}
  \label{tab:noveleval}
  \begin{tabular}{lllll}
    \toprule
    \textbf{Model} & \textbf{Number of Entity Types} & \textbf{Precision} & \textbf{Recall} & \textbf{F1 Score}\\
    \midrule
    Weerawardhana et al.~\cite{10.1007/978-3-319-17040-4_24} & 4 & 0.730 & 0.820 & 0.720 \\
    Li et al.~\cite{9023758} & 4 & 0.839 & 0.789 & 0.813 \\
    Zhou (2022) et al.~\cite{Zhou2022} & 5 & 0.785 & 0.697 & 0.739 \\
    Zhou (2023) et al.~\cite{zhou2023cdtier} & 5 & 0.768 & 0.792 & 0.795 \\
    Ranade et al.~\cite{9671824} & 6 & 0.879 & 0.874 & 0.883 \\
    Wang et al.~\cite{wang2022cyber} & 8 & 0.859 & 0.863  & 0.861 \\
    \textbf{STIXnet Entity Extraction} & \textbf{18} & \textbf{0.903} & \textbf{0.935} & \textbf{0.916} \\
    \bottomrule
\end{tabular}
\end{table*}

We also show the class-specific results for the most frequent entity types in Table~\ref{tab:classpecific}.
As we can see, the \texttt{location} class has the highest performance, given the limited number of entities and their unambiguous nature. Among the best-performing entities, there are also \texttt{intrusion set} and \texttt{malware}, thanks to their peculiar names that are easily identifiable in the text. Classes \texttt{tool} and \texttt{campaign}, however, can cause many false negatives, which is reflected by their recall values. Indeed, these concepts are fairly easy to identify once introduced in the Knowledge Base, but their novel identification might be more difficult. For example, campaign concepts might be hard to recognize in a text, and new tools might be referenced without their introduction (or can be confused with malwares).

\begin{table}[!htpb]
  \centering
  \caption{Entity extraction results for the most frequent STIX entity types.}
  \label{tab:classpecific}
  \begin{tabular}{llll}
    \toprule
    {\bfseries Entity Type} & {\bfseries Precision} & {\bfseries Recall} & {\bfseries F1 Score}\\
    \midrule
    Attack Pattern & 0.702 & 0.861 & 0.771 \\
    Campaign & 0.615 & 0.322 & 0.409 \\
    Identity & 0.719 & 0.878 & 0.790 \\
    Intrusion Set & 0.948 & 0.936 & 0.941 \\
    Location & 0.962 & 0.913 & 0.936 \\
    Malware & 0.888 & 0.790 & 0.835 \\
    Tool & 0.949 & 0.560 & 0.698 \\
    \bottomrule
\end{tabular}
\end{table}

\subsubsection{Temporal Evolution}
\label{subsub:temporal}

The interaction of the entity extraction module with the Knowledge Base allows STIXnet to easily and quickly extract previously recognized entities and constantly update the contents of the database to guarantee high-performance values over time. To ensure that the reports' processing order does not influence the evaluation results, we randomly shuffled the dataset many times and evaluated each shuffle. We obtained a standard deviation between results close to 0 for all the evaluation metrics.

To highlight the capability of STIXnet to maintain its performance over time, we must simulate the execution of the module on subsequent reports following a temporal evolution. To do this, we evaluate its performance according to the following steps:

\begin{enumerate}
    \item We divided the dataset into batches of 5 reports each.
    \item For one of the batches, we run the entire module on each report and evaluate its performance.
    \item Entities found by the novel entity extraction sub-modules are added to the Knowledge Base after a quick manual validation.
    \item Repeat steps (2) and (3) for all the other batches.
\end{enumerate}

\begin{figure*}[!htpb]
  \centering
  \begin{subfigure}{0.32\textwidth}
     \centering
     \includegraphics[width=\textwidth]{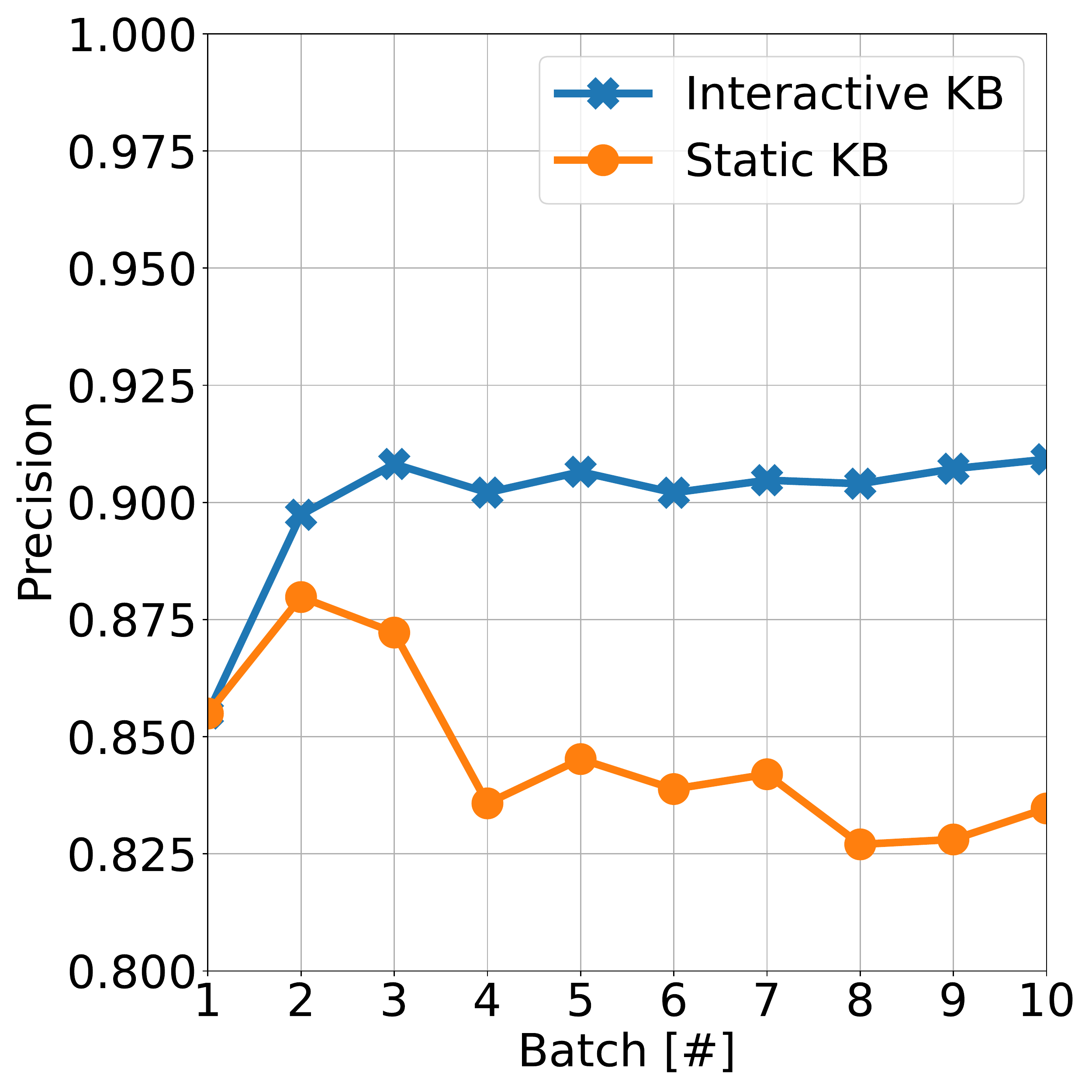}
     \caption{Precision.}
     \label{fig:entextprec}
  \end{subfigure}
  \begin{subfigure}{0.32\textwidth}
     \centering
     \includegraphics[width=\textwidth]{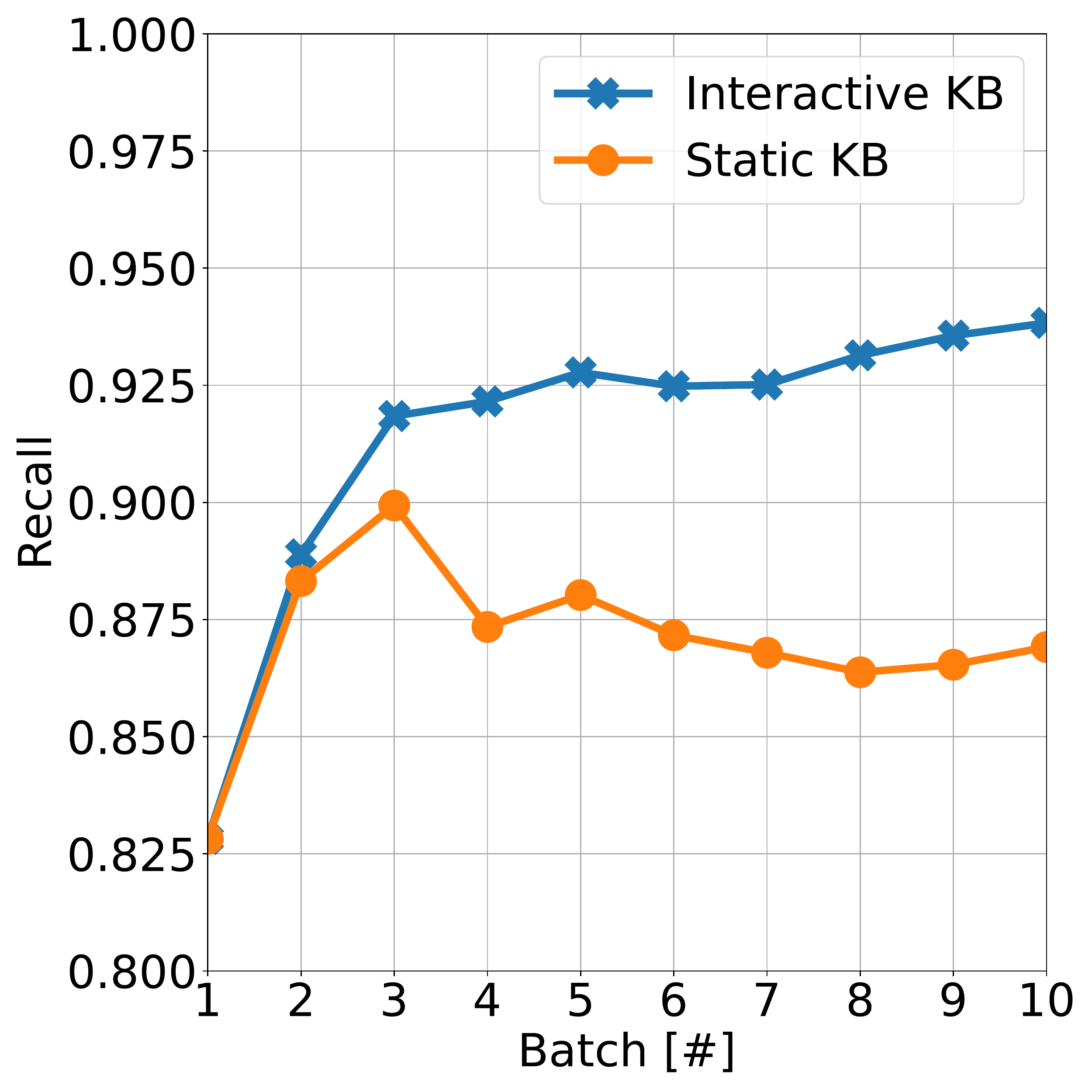}
     \caption{Recall.}
     \label{fig:entextrec}
  \end{subfigure}
  \begin{subfigure}{0.32\textwidth}
     \centering
     \includegraphics[width=\textwidth]{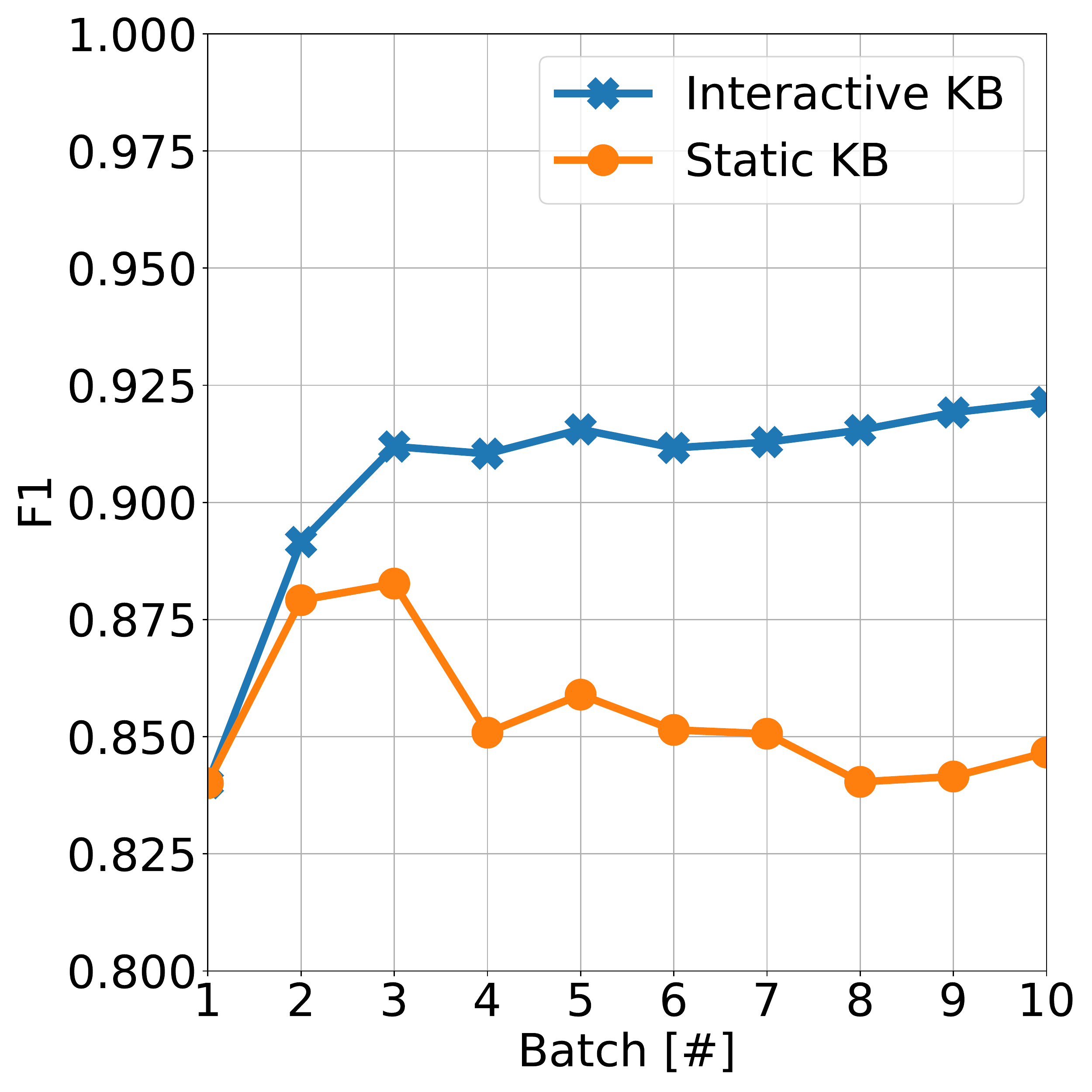}
     \caption{F1 Score.}
     \label{fig:entextf1}
  \end{subfigure}
  \caption{Evolution of the metrics for evaluating Entity Extraction in function of reports batch.}
  \label{fig:metricevol}
\end{figure*}

We think this evaluation is fair on the premises of the possible implementations of the tool. Indeed, in a real-world application, several reports are be published each day and thus need to be processed. Given the deep relationship with the Knowledge Base, we need to avoid introducing bad entities that could degrade the system's performance in the long term. Thus, the newly extracted entities can be manually validated by an analyst, who should ensure the integrity of the data instead of fully annotating the report. The results of this evaluation are shown in Figure~\ref{fig:metricevol}, where the evolution of the metrics is given in function of the processed batch of reports. Furthermore, to highlight the advantage of this approach concerning a static deployment of the Knowledge Base, we performed a different evaluation by repeating the same steps but skipping the third and thus freezing the state of the Knowledge Base in time. As shown, adding entities in the Knowledge Base and their quick validation greatly improve overall performances for all three evaluation metrics and ensure the system's updated status on the latest reports.

\subsection{Relation Extraction}
\label{subsub:relationextraction_results}

For evaluating the relation extraction module of STIXnet, we use the final results derived from the combined use of the rule-based approach and the deep learning based approach for relation extraction. As mentioned, relations are extracted by both submodules with a confidence value which is used to compare the results and eventually impose a threshold of confidence under which relations are discarded. In this way, we include the possibility of a non-relation between two entities. To compare the confidence values for both submodules, we normalized the cosine similarity of the embeddings given by the deep learning based approach in the $[0,1]$ range. We found that a value of 0.5 for the threshold provides a fair tradeoff between false positives and false negatives.

Evaluation has been performed by comparing the relations extracted manually from the reports with the ones that the relation extraction module of STIXnet has found. This module input is constituted by the entities extracted by the entity extraction module and works on them and the sentences in the text to find possible relations. However, the modularity enforced by the STIXnet pipeline introduces the error propagation problem from entity extraction to relation extraction. Indeed, whenever an entity is misclassified by one of the sub-modules of entity extraction, it is passed as input in the relation extraction module, which inevitably produces an error since that entity constitutes a false positive. For this reason, Table~\ref{tab:relext} shows the evaluation for both scenarios. In the first one, we execute the relation extraction module after the results from the entity extraction module have been generated. In the second one, we do not consider relationships between entities where at least one was misclassified to remove the error propagation between the modules. While the F1 scores of the two scenarios are similar, the precision when removing the error propagation effect increases by around 0.1. This type of scenario, however, affects the recall value since the overall number of true positives is decreased, but the number of false negatives is not.

\begin{table}[!htpb]
  \centering
  \caption{Relation Extraction Evaluation.}
  \label{tab:relext}
  \begin{tabular}{llll}
    \toprule
    \textbf{Scenario} & \textbf{Precision} & \textbf{Recall} & \textbf{F1 Score}\\
    \midrule
    Standard & 0.721 & 0.753 & 0.724\\
    No Error Propagation & 0.828 & 0.692 & 0.733 \\
    \bottomrule
\end{tabular}
\end{table}

While the performances of the relation extraction module in the regular scenario have a lower value with respect to the ones of the entity extraction module, we must keep in mind that with each extracted entity, the overall number of possible relationships in the text exponentially increases. Indeed, to the best of our knowledge, ours is the only model that tackles both tasks subsequently by considering each of the entity types of the STIX standard and each of the STIX relationship objects.
\section{Conclusions}
\label{sec:conclusions}

Extracting entities and relations from CTI reports becomes more challenging as the number of classes increases. This paper presents STIXnet, the first solution for automatically extracting all STIX entities and relationships in unstructured Cyber Threat Intelligence reports. Our contribution uses rule-based, NLP, and DL techniques to retrieve all STIX entities and relationships in a report automatically. We also resort to regular expression rules and an extensible Knowledge Base to effectively create an infrastructure for cyber threat analysts to consult and gather all their data and resources. The proposed pipeline enforces a modular approach to be more flexible on the user and their demands, thus separating the different tasks into different modules. The formal definition of the interaction between the sub-modules allows researchers and organizations to use our framework by adding, swapping, and removing sub-modules at will. This is particularly useful in specific scenarios where threat analysts might want to focus on specific entity classes, thus fine-tuning the framework to their needs.

In future works, we would like to explore the text extraction module of STIXnet further to enhance its results. Indeed, being the first module in the pipeline, accurate extraction and efficient artifact removal could improve the precision of the subsequent modules. Furthermore, STIX Domain Objects have many fields that vary depending on the entity types we are considering. All these fields can be filled with information that can be extracted from the sentences. Thus, it could be possible to extend the scope of our processing and provide additional intelligence when present.

\balance
\bibliographystyle{ACM-Reference-Format}
\bibliography{references}

\end{document}